
\nopagenumbers
\magnification=\magstep 1
\vsize=8.5truein
\hsize=6truein
\font\tenrm=cmr10 at 10truept
\overfullrule=0pt
\baselineskip=12pt
\vskip .1in
\centerline{\bf K\"AHLER-CHERN-SIMONS THEORY}
\vskip .3in
\centerline{V.P.NAIR}
\centerline{\it Physics Department, Columbia University}
\centerline{\it New York, NY 10027}
\vskip .3in
\centerline{ABSTRACT}
\parindent=3pc
\baselineskip =10pt
\midinsert
\narrower
{\tenrm K\"ahler-Chern-Simons theory describes antiself-dual gauge fields on a
four-dimensional K\"ahler manifold. The phase space is the
space of gauge potentials, the symplectic reduction of which by the
constraints of antiself-duality leads to the moduli space of
antiself-dual instantons. We outline the theory highlighting
symmetries, their canonical
realization and some properties of the quantum wave functions.
The relationship to integrable systems {\it via} dimensional
reduction is briefly
discussed.}
\endinsert
\baselineskip =12pt
\parindent=2pc
\vskip .3in
In this talk, I shall describe some recent work done in
collaboration with Jeremy Schiff on what we refer to as
K\"ahler-Chern-Simons (KCS) theory.$^1$ The theory basically provides an
action description of antiself-dual gauge fields, i.e.
instantons on four-dimensional K\"ahler manifolds. The motivation
for seeking such a theory is essentially twofold. There is
considerable evidence that antiself-dual gauge fields may be
considered as a `master' integrable system.$^2$ For example, we can
consider ${\bf R}^4$ as a K\"ahler manifold, pairing up the standard
coordinates into complex ones as $z=x_{2}+ix_{1},~~w=x_{4}+ix_{3}$.
The conditions of antiself-duality are then given by
$$
F_{zw}~=F_{{\bar z}{\bar w}}~=F_{z{\bar z}}+F_{w{\bar w}}~=~0
\eqno(1)
$$
where $F_{ab}$ denotes the $(ab)$ component of the field strength,
which is as usual valued in the algebra of a Lie group $G$.
Dimensional reduction of these equations, such as the requirement
of the fields being independent of, say ${\bar w}$, gives a large
class of known two-dimensional integrable theories such as the
Korteweg-de Vries (KdV),
nonlinear Schr\"odinger, Boussinesq and other equations.$^{3,4}$ It is
likely that all integrable theories can be considered as special
cases of the antiself-dual gauge theory. Now, integrable theories themselves
are of interest because they may describe certain types of
perturbations of conformal field theories and
also because the Poisson bracket structures associated with
integrable theories are related to the Virasoro and $W_{N}$ algebras.$^5$
These algebras in turn can be the chiral algebras of conformal field
theories. Independently, the study of gravity theories associated to
these algebras
also seem to lead to four-dimensional K\"ahler manifolds.$^6$
A unified description of integrable theories in terms of antiself-dual
gauge fields, especially in a Lagrangian framework with associated
Hamiltonian and Poisson bracket structures, can thus be useful in
understanding these theories and algebras.

Another way to introduce KCS theory would be as a
generalization of Chern-Simons theory in $2+1$ dimensions.
Let me explain by recalling that Chern-Simons theory is described by the
action
$$
S~=~ -{k\over {4\pi}} \int_{\Sigma\times {\bf R}}~Tr\bigl(
AdA+{\textstyle {2\over 3}}A^{3}\bigr)\eqno(2)
$$
In the Hamiltonian quantization of this theory that we want to focus on,
$\Sigma$ is in general a Riemann surface and the coordinate representing
$\bf R$ will be taken as time. The wave functionals obtained upon
quantization of this theory are the chiral blocks of a conformal field
theory (specified by choice of $k$ and gauge group $G$) on $\Sigma$. Thus
they carry a representation of the holomorphic current algebra.$^7$ A natural
higher dimensional generalization would be a four-dimensional
gauge theory with a holomorphic symmetry algebra.

At first glance, these two motivations seem somewhat disjoint, but it is
easy to see that the same theory would be the result. The phase space
of the Chern-Simons theory is given by the Wilson lines or holonomies
associated with the noncontractible paths in $\Sigma$, i.e. by the
homomorphisms $\pi_{1}(\Sigma)\rightarrow G$.
This space can also be identified as the moduli space of (stable) holomorphic
vector bundles of rank $N$ on $\Sigma$ (for $G=SU(N)$).$^8$
The four-dimensional generalization of this
result is that the moduli space of (stable) holomorphic vector bundles of
rank $N$ on a
compact K\"ahler manifold $M$ is essentially the moduli space of
(irreducible) antiself-dual
gauge
fields on $M$ (for $G=SU(N)$).$^9$
Thus a generalization of Chern-Simons theory in terms of the nature of its
phase
space would naturally lead us to antiself-dual gauge fields.

The notion of self-duality or antiself-duality is also crucial in the
dynamics of $N=2$ strings.$^{10}$ For gauge fields, the most appropriate theory
would be $N=2$ heterotic strings. The KCS theory is an effective
Lagrangian description, in terms of fields on the target space, of
this theory. (Of course, this is for the case when there are only gauge
fields; our description of the target space dynamics
has to be augmented when gravitational
excitations are also present.)

The action for the KCS theory can be written as
$$
S~=~ \int_{M\times {\bf R}}-{k\over {4\pi}}Tr
\bigl(AdA+{\textstyle {2\over 3}}A^{3}\bigr) \omega~+~ Tr \bigl(\Phi{\cal
F}+{\bar \Phi} {\cal F}\bigr)\eqno(3)
$$
where ${\cal F}$ is the field strength on $M\times {\bf R}$, $F$ will
denote the field strength on $M$ when needed. $M$ is a K\"ahler manifold
of real dimension four, $\omega$ is the K\"ahler form, related as usual
to the metric by $ds^{2}= g_{a{\bar a}}dz^{a}d{\bar z}^{\bar a},~
\omega= {i\over 2} g_{a{\bar a}} dz^{a}\wedge d{\bar z}^{\bar
a}~=i\partial {\bar \partial}K$, where $z^{a}, a=1,2$ are local complex
coordinates on $M$ and $K$ is the K\"ahler potential.
Exterior products of the differential forms in (3) are understood.
$\Phi$ is a Lie algebra valued (2,0) form on $M$ and a one-form
on $\bf R$; i.e. in local coordinates
$$
\Phi= \phi dt~= {\textstyle {1\over 2}}\phi_{ab}dz^{a}dz^{b}dt\eqno(4)
$$
The fields have the standard gauge transformation properties.
$$
A^{u}= uAu^{-1}-du~u^{-1},~~~~~~~~~~~ \Phi^{u}=u\Phi
u^{-1}\eqno(5)
$$
where $u$ is a locally defined $G$-valued function on $M\times {\bf
R}$. Since $\omega$ is a closed two-form, the action (3) is indeed
invariant under gauge transformations which are homotopic to the
identity. Invariance under gauge transformations which are homotopically
nontrivial, if they exist, will require $k$ to be an integer.

The equations of motion for the action (3) are
$$
F^{(0,2)}~=F^{(2,0)}~=0\eqno(6a)
$$
$$
F\wedge \omega~=0\eqno(6b)
$$
$$
{k\over {4\pi}}{\dot A}_{a}~= -ig^{{\bar a}b} {\overline \nabla}_{\bar
a}\phi_{ba}\eqno(6c)
$$
$$
{k\over {4\pi}}{\dot A}_{\bar a}~=~ig^{{\bar b}a}\nabla_{a}{\bar \phi}_{{\bar
b}{\bar a}}\eqno(6d)
$$
Here $\nabla, ~{\overline \nabla}$ denote the gauge and Levi-Civita
covariant derivatives and ${\dot A}$ denotes the time derivative of $A$;
we have used the $A_{t}=0$ gauge in (6).
Equations (6a,b) follow from variation with respect to $\Phi,~{\bar
\Phi}$ and the time component of the gauge potential $A_{t}$. They are
constraints on the initial data for gauge fields, i.e. the fields on
$M$, requiring them to be antiself-dual. The time evolution equations
(6c,d) follow from variation with respect to the space ($M$-)components of the
potential. Since the constraints (6a,b) hold for all time, we have
$ {\dot F}^{(0,2)}=0$; this leads to the gauge covariant
Laplace equation for $\phi$
$$
\nabla \ast {\overline \nabla}\ast \phi ~=0\eqno(7)
$$
Here $\ast$ denotes the Hodge star operation.
On manifolds with $\partial M=\emptyset$, using a Bochner-type argument,
one can show that there is no
solution for $\phi$, at least for manifolds of positive scalar
curvature.$^{11}$ Thus ${\dot A}=0$.
Time evolution is trivial and the set of classical solutions is given by
antiself-dual gauge fields on $M$. For a general $M$, there can be
nontrivial solutions for $\phi$; it will turn out that these can be
related to B\"acklund transformations. In any case, the Hamiltonian is a
sum of the constraints $F^{(0,2)},~F^{(2,0)}$ and $F\wedge \omega$ and
therefore any time evolution is a change of `gauge' in the sense of
Dirac's theory of constraints; it is thus effectively trivial.

The KCS theory (3) thus provides an action description of antiself-dual
gauge
fields. The Hamiltonian structure of the theory is as follows. The phase
space is the space ${\cal A}$ of gauge potentials on $M$.
The symplectic two-form on this space can be easily read off from the
action since the latter is first order in time derivatives and is given
by
$$
\Omega~=~ {k\over {4\pi}} \int_{M} Tr \bigl( \delta A~\delta
A\bigr)\omega ~=~{ik\over {2\pi}}\int_{M} dV~g^{{\bar a}a}\delta A^{i}_{\bar a}
\delta A^{i}_{a}\eqno(8)
$$
where $\delta$ denotes exterior differentiation on ${\cal A}$ and $dV$
is the volume element on $M$. The superscript on the potentials
refers to the component with respect to a chosen basis for the Lie
algebra of $G$.
The cohomology class of $\Omega$ is
unchanged by gauge transformations of $A$ or by the addition of exact
terms to $\omega$. It
is an example of the Donaldson map $H^{2}(M)\rightarrow H^{2}({\cal
A}/{\cal G})$ on the cohomology groups,$^{12}$ where ${\cal G}$ denotes the
group of gauge transformations.
The Poisson brackets corresponding to (8) are given by
$$
\bigl[ A^{i}_{a}(x),~A^{j}_{\bar a}(y)\bigr]~=~{{2\pi}\over
{ik}}g_{a{\bar a}} \delta^{ij} ~{{\delta^{(4)}(x-y)}\over {{\rm det}(g)}}
\eqno(9)
$$

One has to further take care of the constraints (6a,b) on ${\cal A}$.
This of course involves the symplectic or Hamitonian reduction of ${\cal
A}$ essentially by the constraints $F^{(0,2)}$ and $F\wedge \omega $.
The reduced phase space is the moduli space of antiself-dual
gauge fields on $M$.

For the remainder of this talk I shall discuss the constraints and the
reduction of phase space and some aspects of the quantum theory and
dimensional reduction on ${\bf R}^4$. The algebra of the constraints
is a necessary prerequisite for proper reduction of the phase space.
It turns out that
$F\wedge \omega $ and either of the other two, say $F^{(0,2)}$,
are the relevant first class
constraints. $F\wedge \omega $ is the generator of gauge
transformations; $F^{(0,2)}$ generates infinitesimal B\"acklund
transformations in those cases where the Laplace equation (7) has
nontrivial solutions.
The wave functionals in the quantum theory have a factor $e^{i{\cal S}[U]}$
where $U$ is a locally defined function taking values in $G^{\bf C}$, the
complexification of $G$.
${\cal S}[U]$ is a generalization of the Wess-Zumino-Witten (WZW) action
in two dimensions$^{13}$ and obeys an analogous Polyakov-Wiegmann factorization
property. This property leads to holomorphic and antiholomorphic
symmetries for ${\cal S}[U]$ and associated current algebras. Finally we
shall consider dimensional reduction of the conditions of
antiself-duality for the case of $M$ being ${\bf R}^4$. In the framework of
reduction of ${\cal A}$ by $F^{(0,2)}$ followed by $F\wedge \omega$, we
shall see that many of the different ans\"atze which have been used by
various authors are indeed obtained by suitable gauge choices and are
related to each other.

A remark before we turn to the details: notice that there is considerable
similarity between the present discussion and $2+1$ dimensional
Chern-Simons theory. In the latter case, the holomorphic and
antiholomorphic components of the gauge potential, {\it viz}. $A_{z}$ and
$A_{\bar z}$ are canonically conjugate to each other, analogous to (9).
The wave functionals likewise have a factor $e^{i{\cal S}[U]}$ where
${\cal S}[U]$ is the WZW action. The holomorphic and antiholomorphic
current algebras are, of course, the Kac-Moody algebras.$^7$

We now turn to the reduction of ${\cal A}$. The first step is the
algebra of constraints. We introduce test functions
$\varphi,~{\bar\varphi}$ and $\theta$ which take values in the Lie
algebra of $G$ and serve as parameters for the transformations generated
by $F^{(0,2)},~F^{(2,0)}$ and $F\wedge \omega$ respectively.
${\bar \varphi}$ is a $(0,2)$ form and $\theta$ is a
scalar. The generators are collected as
$$
E({\bar\varphi})=-{k\over {2\pi}}\int_{M}Tr ({\bar\varphi}F)~~~~~~~~~
{\overline E}(\varphi )=-{k\over {2\pi}}\int_{M}Tr (\varphi F) \eqno(10)
$$
$$
G(\theta)= -{k\over {2\pi}}\int_{M} Tr (\theta F) \eqno(11)
$$
The Poisson brackets of these generators and the potential $A$ are given
by
$$
[G(\theta),A^{i}_{a}(x)]=-(\nabla \theta)^{i}_{a}(x)~~~~~~~~~
[G(\theta), A^{i}_{\bar a}(x)]=-({\overline \nabla}\theta)^{i}_{\bar a}(x)
$$
$$
[E({\bar \varphi}),
A^{i}_{a}(x)]=0~~~~~~~~~~~~~~~~~[E({\bar\varphi}),A^{i}_{\bar a}(x)]=
{}~i(\ast \nabla \ast {\bar\varphi} )^{i}_{\bar a}(x)
$$
$$
[{\overline E}(\varphi),A^{i}_{\bar a}(x)]=0~~~~~~~~~~~~~~~~~
[{\overline E}(\varphi),A^{i}_{a}(x)]=-i(\ast {\overline \nabla}\ast
\varphi)^{i}_{a}(x)
\eqno(12)
$$
\vskip .1in
$$
[G(\theta),G(\theta ')]=G(\theta\times\theta ')
$$
$$
[G(\theta), E({\bar \varphi})]=E(\theta \times {\bar\varphi})
$$
$$
[G(\theta),{\overline E}(\varphi)]={\overline E}(\theta\times\varphi)
$$
$$
[{\overline E}(\varphi), E({\bar\varphi}')]~={ik\over {2\pi}}\int_{M}
Tr \bigl({\bar\varphi}' \nabla \ast {\overline \nabla}\ast \varphi \bigr)
\eqno(13)
$$
Here the cross product is in the Lie algebra, i.e. ${\theta\times
\theta '}^{i}
= f^{ijk} \theta^{j}\theta '^{k}$, $f^{ijk}$ being the structure
constants of the Lie algebra.
The first set of equations show that $G(\theta)$ is the generator of
gauge transformations and that ${\overline E}$ and $E$ generate changes in
the potentials, respectively, of the form
$$
A_{a}\rightarrow A_{a}~-i(\ast {\overline \nabla}\ast \varphi )_{a}
$$
$$
A_{\bar a}\rightarrow A_{\bar a}~+i(\ast \nabla \ast {\bar
\varphi})_{\bar a}\eqno(14)
$$

Consider now the reduction of ${\cal A}$. We start by setting ${\overline E}$
or $F^{(0,2)}$ to zero. The solution set for this condition is given by
$$
A=\bigl( A_{a},~-\partial_{\bar a}UU^{-1}\bigr)\eqno(15)
$$
where $U$ is a locally defined $G^{\bf C}$-valued function.
If the equation
$$
\nabla \ast {\overline \nabla}\ast \varphi =0\eqno(16)
$$
has no
solution, then ${\overline E}$ and $E$ have nonvanishing Poisson brackets,
as seen from (13), and
$E=0$ can be used as gauge fixing condition for the flow generated by
${\overline E}$. The reduced phase space is characterized by the vanishing of
${\overline E}$ and $E$. Further reduction is achieved by setting $G(\theta)$
to zero; this is equivalent to
$$
g^{a{\bar a}}\partial_{\bar a}\bigl(J^{-1}\partial_{a}J\bigr)=0\eqno(17)
$$
where $J=U^{\dagger}U$. The reduced fields are antiself-dual
gauge fields and the
reduced phase space is the moduli space of antiself-dual fields.

Consider now the case when (16)
has nontrivial solutions. The flow generated by ${\overline E}$ on the
${\overline
E}=0$ subspace is given by
$$
U\rightarrow U
$$
$$
A_{a}\rightarrow A_{a}'=A_{a}~-i \bigl( \ast {\overline \nabla}\ast
\varphi\bigr)_{a}\eqno(18)
$$
For those $\varphi$ which are solutions to (16), we see that both
$A_{a}$ and $A_{a}'$ satisfy the condition $E=0$. Writing $\ast {\overline
\nabla}\ast \varphi~=\nabla \sigma$, we get $A_{a}=(U^{\dagger
-1}\partial_{a}U^{\dagger}),~ A_{a}'= (U'^{\dagger
-1}\partial_{a}U'^{\dagger})$, where
$U'^{\dagger}=U^{\dagger}e^{-i\sigma}$. We can rewrite these as
$$
(J^{-1}\partial_{a}J)~-~ (J'^{-1}\partial_{a}J')~=~ i(U^{-1}\nabla
\sigma ~U)\eqno(19)
$$
We can check easily that if $J$ satisfies (17), then $J'$ defined
by the solution of the first order equation (19) also satisfies (17).
Thus the transformations generated by ${\overline E}$, with parameters which
are solutions to (16), generate solutions to the antiself-dual conditions from
solutions; i.e. it is an infinitesimal B\"acklund transformation.
It is important that these are now canonically realized.

In the case of ${\bf R}^4$, it is possible to choose
$\varphi_{ab}={1\over 2} \lambda \epsilon_{ab}\sigma$; equation (19)
then becomes
$$
(J^{-1}\partial_{a}J)~-~(J'^{-1}\partial_{a}J')~=~-\lambda~ g^{{\bar a}b}
\epsilon_{ba}\partial_{\bar a}(J^{-1}J')\eqno(20)
$$
This is the more conventional way of writing B\"acklund
transformations.$^{14}$

As regards the quantum theory, we shall concentrate on the question of
holomorphic current algebra. Since $A_{a}$ and $A_{\bar a}$ are
canonically conjugate, we can take the wave functions to be functionals
of only one of these sets, say $A_{\bar
a}$. Once we require ${\overline E}=0$, i.e. with $A_{\bar a}=-\partial_{\bar
a}U~U^{-1}$, we can take the wave functions to be functionals of $U$.
The inner product is then given by
$$
\big<\Psi_{1}\vert \Psi_{2}\big>~=~ \int [dU] ~e^{-{\tilde
K}}\Psi_{1}^{*}\Psi_{2}\eqno(21)
$$
where ${\tilde K}$ is the K\"ahler potential associated with $\Omega$,
${\tilde K}= {k\over {2\pi}} \int dV~ A^{i}_{a}A^{i}_{\bar a}g^{a{\bar a}}$.
For the purposes of discussing the current algebra, it suffices to
consider the case of zero instanton number, that is, the case for which
the second Chern class of
the gauge fields on $M$ is trivial; in this case $U$ can be globally
defined. The second stage of reduction is performed by requiring
$$
G(\theta)~\Psi [U]=0\eqno(22)
$$
The solution to this equation is given by $\Psi [U]=e^{i{\cal S}[U]}$,
where
$$
{\cal S}[U]~= {k\over {2\pi}}\int_{M} dV~ g^{{\bar a}a}
Tr\bigl(\partial_{a}U~\partial_{\bar a}U^{-1}\bigr)~+~ {ik\over {12\pi}}
\int_{M^5}Tr \bigl(U^{-1}dU\bigr)^{3}\omega \eqno(23)
$$
$M^5$ is taken to be $M\times [0,1]$; one component of the boundary of
$M^5$ is identified as our space $M$. The field $U$ which is defined on $M$ is
extended to $M^5$ in such a way that it goes to a fixed function $U_0$
on the other component, same for all $U$ within the same homotopy class.
${\cal S}[U]$ is the analogue of the WZW action in two dimensions.$^{13}$
It obeys a Polyakov-Wiegmann type factorization formula
$$
{\cal S}[U_{1}U_{2}]={\cal S}[U_{1}]~+~ {\cal S}[U_{2}]~-~{k\over {\pi}}
\int_{M}dV ~g^{{\bar a}a} Tr \bigl(
U_{1}^{-1}(\partial_{a}U_{1})(\partial_{\bar
a}U_{2})U_{2}^{-1}\bigr)\eqno(24)
$$
This formula shows that transformations on $U$ of the form $U\rightarrow
h~U~{\tilde h}$, where $h$ is antiholomorphic and ${\tilde h}$ is
holomorphic, is a symmetry of ${\cal S}$.
As in the case of WZW action in two dimensions, one can think of ${\cal
S}$ as the starting point, choosing, say
$z_{1}$ as the time coordinate and obtain the generators and algebra of
these symmetries. The algebra one obtains, for the generators $Q^{i}({\bar
z})$ of the antiholomorphic
transformations, is
$$
[Q^{i}({\bar z}), Q^{j}({\bar z}')]~= f^{ijk}Q^{k}({\bar
z})\delta^{2}({\bar z}-{\bar z}')-{k\over {4\pi}}\delta^{ij}C({\bar
z},{\bar z}')
$$
$$
C({\bar z},{\bar z}')=~\int ~dz_{2}~{\rm det}(g) g^{{\bar a}1}
\partial_{\bar a}\delta^{(2)}({\bar z}-{\bar z}')\eqno(25)
$$
This algebra, despite its similarity to the Kac-Moody algebra, is very
limited in its utility in solving the theory defined by (23), since,
unlike its two-dimensional counterpart,
the classical solutions do not factorize into holomorphic and
antiholomorphic matrices.

We now turn to dimensional reduction on ${\bf R}^4$. Our remarks earlier
concerning gauge choices can be illustrated by considering $G=SL(2,{\bf
C})$. Starting with the four coordinates $z,{\bar z}, w$ and ${\bar w}$,
we do a dimensional reduction by considering gauge potentials which are
independent of one of them, say ${\bar w}$, in some gauge. This still
leaves us the freedom of gauge transformations which are independent of
${\bar w}$; under these $A_{\bar w}\rightarrow uA_{\bar w}u^{-1}$, where
$u$ is the gauge transformation matrix which now depends only on $z{,\bar
z},w$.
Potentials $A_{\bar w}$ fall into two classes, characterized by the following
canonical forms, to which they can be brought by use of the above gauge
freedom.
$$
A_{\bar w}=\left(\matrix{0&0\cr
1&0\cr}\right) ~~~~or~~~~~\kappa\left(\matrix{1&0\cr
0&-1\cr}\right) \eqno(26)
$$
The first of these leads to the KdV and the modified KdV equations,
the second gives the
nonlinear Schr\"odinger equation. We now impose
$$
F_{{\bar z}{\bar w}}=0~~~~~~~~
F_{z{\bar z}}+F_{w{\bar w}}=0\eqno(27)
$$
These equations can be solved for the other components of the potential;
for example, for the KdV choice
$$
A_{w}=\left(\matrix{(j_{z}-f_{\bar z}-2dj)/2&d_{\bar z}-j\cr
c&-(j_{z}-f_{\bar z}-2dj)/2\cr}\right)
$$
$$
A_{z}=\left(\matrix {d&1\cr
f&-d\cr}\right)~~~~~~~~A_{\bar z}=\left(\matrix{0&0\cr
j&0\cr}\right)\eqno(28)
$$
$c,d,f,j$ are arbitrary functions of $z,{\bar z},w$ and the subscripts
denote differentiation with respect to the coordinates indicated.
There is some freedom in choosing a solution to (27) due to integration
constants, etc; we have made a specific choice. There is also some gauge
freedom left; thus for the KdV choice of $A_{\bar w}$, we can still do
gauge transformations by matrices of the form
$$
u=\left(\matrix{1&0\cr \gamma &1\cr}\right)\eqno(29)
$$
By a suitable choice of $\gamma$ we can set $j$ to zero whereupon the
last of the antiself-duality conditions, $viz$. $F_{zw}=0$ becomes the KdV
equation. This is the choice made by Mason and Sparling.$^3$ Another choice,
made by Bakas and Depireux,$^4$
is to set $d$ to zero; this also leads to the KdV equation. A third
possibility, $viz$. choosing $f$ to be zero gives the modified KdV
equation.  Notice that the various ans\"atze for the potentials are not
{\it ad hoc} choices for us; the gauge freedom naturally leads us to
these. The relationships among the various choices is also clear.
\vskip .2in
\noindent
{\bf References}
\vskip .1in
\item
{1.} V.P.Nair and J.Schiff, {\it Phys.Lett.} {\bf B246} (1990) 423; {\it
K\"ahler-Chern-Simons Theory and Symmetries of Antiself-Dual Gauge
Fields}, Columbia Preprint CU-TP-521 (1991).
\vskip .1in
\item
{2.} R.S.Ward, {\it Phil.Trans.Roy.Soc.Lond.} {\bf A315} (1985) 451.
\vskip .1in
\item
{3.} L.J.Mason and G.A.J.Sparling, {\it Phys.Lett.} {\bf A137} (1989) 29.
\vskip .1in
\item
{4.} I.Bakas and D.Depireux, {\it Mod.Phys.Lett.} {\bf A6} (1991) 399;
Maryland Preprints UMD-PP91-111,168.
\vskip .1in
\item
{5.} J.L.Gervais, {\it Phys.Lett.} {\bf B160} (1985) 277; for further
reference, see P.di Francesco, C.Itzykson and J.B.Zuber, Santa Barbara,
Princeton and Saclay Preprint NSF-ITP-90-193, PUPT-1211, SPhT/90-149.
\vskip .1in
\item
{6.} K.Schoutens. A.Sevrin and P.van Nieuwenhuizen, Stony Brook Preprints
ITP-SB-90-19,39,62.
\vskip .1in
\item
{7.} E.Witten, {\it Comm.Math.Phys.} {\bf 121} (1989) 351; M.Bos and
V.P.Nair, {\it Phys.Lett.} {\bf B223} (1989) 61; {\it Int.J.Mod.Phys.}
{\bf A5} (1990) 959; S.Elitzur {\it et al, Nucl.Phys.} {\bf B 326}
(1989) 108; J.M.F.Labastida and A.V.Ramallo, {\it Phys.Lett.} {\bf B227}
(1989) 92; H.Murayama, {\it Z.Phys.} {\bf C48} (1990) 79;
T.Killingback, {\it Phys.Lett.} {\bf B219} (1989) 448;
A.Polychronakos, {\it Ann.Phys.} {\bf 203} (1990) 231;
Florida Preprint UFIFT-HEP-89-9; T.R.Ramadas, I.M.Singer and J.Weitsman,
MIT Mathematics Preprint (1989).
\vskip .1in
\item
{8.} M.F.Atiyah and R.Bott, {\it Phil.Trans.Roy.Soc.Lond.} {\bf A308}
(1982) 523.
\vskip .1in
\item
{9.} S.K.Donaldson, {\it Proc.Lond.Math.Soc.}(3) {\bf 50} (1985) 1.
\vskip .1in
\item
{10.} H.Ooguri and C.Vafa, Harvard Preprints HUTP-91/A003 and A004.
\vskip .1in
\item
{11.} M.Itoh, {\it Publ. RIMS, Kyoto Univ.} {\bf 19} (1983) 15;
{\it J.Math.Soc.Japan} {\bf 40} (1988) 9.
\vskip .1in
\item
{12.} P.Braam, {\it The Fascinating Relations between 3- and 4-Manifolds
and Gauge Theory,} Lectures at the Fifth Annual University of California
Summer School on Nonlinear Science, {\it Physics and Geometry,} Lake
Tahoe, 1989.
\vskip .1in
\item
{13.} E.Witten, {\it Comm.Math.Phys.} {\bf 92} (1984) 455.
\vskip .1in
\item
{14.} L.Dolan, {\it Phys.Rep.} {\bf 109} (1983) 1; L.L.Chau in {\it
Integrable Systems}, X.C.Song (ed.) (World Scientific, 1990);
L.Crane, {\it Comm.Math.Phys.} {\bf 110} (1987) 391.

\bye